\documentclass[unrestricted]{stagstatus}
\usepackage{hyperref}
\usepackage{blindtext}
\usepackage[most]{tcolorbox}
\usepackage{graphicx}
\usepackage{listings}
\usepackage{booktabs}
\usepackage{subcaption}
\usepackage{import}

\usepackage{amsmath}
\usepackage{amssymb}
\usepackage{amsthm}

\usepackage{wrapfig}
\usepackage{placeins}

\usepackage[simplified]{pgf-umlcd}
\usepackage{tikzit}
\usetikzlibrary{calc}

\tikzstyle{basic}=[fill=white, draw=black, shape=circle]
\tikzstyle{square}=[fill=white, draw=black, shape=rectangle]
\tikzstyle{big dashed}=[fill=white, draw=black, shape=circle, minimum width=1cm, dashed]
\tikzstyle{vertical ellipse dashed}=[fill=none, draw=blue, minimum width=0.75cm, minimum height=3cm, ellipse, dashed, tikzit shape=rectangle, tikzit draw=blue, tikzit fill=white]
\tikzstyle{small vertical ellipse dashed}=[fill=none, draw=blue, shape=circle, tikzit fill=white, tikzit draw=blue, dashed, minimum width=0.75cm, minimum height=1.5cm, tikzit shape=rectangle, ellipse]
\tikzstyle{tiny vertical ellipse dashed}=[fill=none, draw=blue, shape=circle, tikzit fill=white, ellipse, dashed, minimum width=0.75cm, minimum height=1cm, tikzit shape=rectangle]
\tikzstyle{red}=[fill={rgb,255: red,191; green,0; blue,64}, draw=black, shape=circle]
\tikzstyle{green}=[fill={rgb,255: red,0; green,128; blue,128}, draw=black, shape=circle]
\tikzstyle{blue}=[fill=blue, draw=black, shape=circle]
\tikzstyle{huge dashed}=[fill=white, draw=black, shape=circle, dashed, minimum width=2cm]
\tikzstyle{medium}=[fill=white, draw=black, shape=circle, minimum width=1cm]
\tikzstyle{pale green}=[fill={rgb,255: red,173; green,231; blue,0}, draw=black, shape=circle, minimum width=1cm]
\tikzstyle{horizontal ellipse dashed}=[fill=white, draw=black, tikzit draw=magenta, tikzit shape=rectangle, minimum width=3cm, minimum height=0.75cm, ellipse, dashed]
\tikzstyle{minsize}=[fill=white, draw=black, shape=circle, minimum width=0.75cm]
\tikzstyle{horizontal ellipse green}=[fill={rgb,255: red,191; green,255; blue,0}, draw=black, tikzit draw={rgb,255: red,191; green,255; blue,0}, tikzit shape=rectangle, minimum width=3cm, minimum height=0.75cm, ellipse, dashed]
\tikzstyle{horizontal ellipse blue}=[fill={rgb,255: red,107; green,203; blue,255}, draw=black, tikzit draw=blue, tikzit shape=rectangle, minimum width=3cm, minimum height=0.75cm, ellipse, dashed]
\tikzstyle{smallblack}=[fill=black, draw=black, shape=circle, inner sep=0 pt, minimum size=3 pt]
\tikzstyle{smallSquare}=[fill=white, draw=black, shape=rectangle, inner sep=0 pt, minimum size=6 pt]
\tikzstyle{smallCircle}=[fill=white, draw=black, shape=circle, inner sep=0 pt, minimum size=6 pt]
\tikzstyle{big vertical ellipse dashed}=[fill=none, draw=blue, shape=circle, tikzit shape=rectangle, ellipse, dashed, minimum width=0.95cm, minimum height=3.7cm]
\tikzstyle{smallred}=[fill={rgb,255: red,191; green,0; blue,64}, draw=black, shape=circle, inner sep=0 pt, minimum size=6 pt]
\tikzstyle{smallblue}=[fill=blue, draw=blue, shape=circle, inner sep=0pt, minimum size=3pt]

\tikzstyle{directed}=[->, line width=1pt]
\tikzstyle{undirected}=[-, line width=1pt]
\tikzstyle{directed red}=[draw=red, ->, line width=1pt]
\tikzstyle{directed green}=[draw={rgb,255: red,0; green,128; blue,128}, ->, line width=1pt]
\tikzstyle{directed blue}=[draw=blue, ->, line width=1pt]
\tikzstyle{directed purple}=[draw={rgb,255: red,128; green,0; blue,128}, ->, line width=1pt]
\tikzstyle{undirected red}=[-, draw=red, line width=1pt]
\tikzstyle{undirected green}=[-, draw={rgb,255: red,0; green,107; blue,61}, line width=1pt]
\tikzstyle{undirected blue}=[-, draw=blue, line width=1pt]
\tikzstyle{undirected purple}=[-, draw={rgb,255: red,128; green,0; blue,128}, line width=1pt]
\tikzstyle{undirected dashed}=[-, line width=1pt, dashed]
\tikzstyle{orange dashed}=[-, draw={rgb,255: red,255; green,128; blue,0}, dashed, line width=1pt]
\tikzstyle{directed dash}=[->, dashed]
\tikzstyle{blue dashed}=[-, draw=blue, dashed, line width=1pt]
\tikzstyle{green dashed}=[-, draw={rgb,255: red,0; green,162; blue,0}, dashed, line width=1pt]
\tikzstyle{blue filled}=[-, fill={blue!20}, draw=blue, line width=1pt, opacity=0.5, tikzit fill=white]
\tikzstyle{red filled}=[-, fill={red!20}, line width=1pt, draw=red, opacity=0.5, tikzit fill=white]
\tikzstyle{green filled}=[-, line width=1pt, draw={rgb,255: red,0; green,107; blue,61}, opacity=0.5, tikzit fill=white, fill={rgb,255: red,149; green,255; blue,179}]
\tikzstyle{orange filled}=[-, fill={orange!20}, draw=orange, line width=1pt, opacity=0.5, tikzit fill=white]
\tikzstyle{undirected dashed}=[-, draw=black, dashed, line width=1pt]

\renewcommand{\epsilon}{\varepsilon}

\newcommand{\p}{\mathbb{P}}

\newcommand{\R}{\mathbb{R}}

\newcommand{\Z}{\mathbb{Z}}

\definecolor{indiagreen}{rgb}{0.07, 0.53, 0.03}

\newcommand{\stagversion}{2.0}

\usepackage{todonotes}
\setuptodonotes{color=blue!30, size=\tiny}
\definecolor{eggshell}{rgb}{0.94, 0.92, 0.84}

\definecolor{codegreen}{rgb}{0,0.6,0}
\definecolor{codegray}{rgb}{0.5,0.5,0.5}
\definecolor{codepurple}{rgb}{0.58,0,0.82}
\definecolor{backcolour}{rgb}{0.95,0.95,0.92}

\lstdefinestyle{mystyle}{
    backgroundcolor=\color{backcolour},   
    commentstyle=\color{codegreen},
    keywordstyle=\color{magenta},
    numberstyle=\tiny\color{codegray},
    stringstyle=\color{codepurple},
    basicstyle=\ttfamily\footnotesize,
    breakatwhitespace=false,         
    breaklines=true,                 
    captionpos=b,                    
    keepspaces=true,                 
    numbers=left,                    
    numbersep=5pt,                  
    showspaces=false,                
    showstringspaces=false,
    showtabs=false,                  
    tabsize=2
}

\title{Technical Report~(2)}
\statusdate{\today}
\project{EPSRC Early Career Fellowship,  EP/T00729X/1.}

\begin{document}

\frontmatter

\setcounter{page}{1}

\section{Summary}
Spectral Toolkit of Algorithms for Graphs (STAG) is an open-source C++ and Python library providing several methods for working with graphs and performing graph-based data analysis.
In this technical report, we provide an update on the development of the STAG library.
The report serves as a user's guide for the newly implemented algorithms, and gives implementation  details       and engineering choices made in the development of the library. The report is structured as follows: 
\begin{itemize}
\item Section~\ref{sec:lsh} 
 describes the locality sensitive hashing,  and the main components used in its construction.   
\item Section~\ref{sec:kde} 
describes the  kernel density estimation, and the state-of-the-art algorithm for the kernel density estimation.   The
discussion is at a high level, and  domain knowledge beyond basic algorithms is not needed.
\item Section~\ref{sec:sc} describes a fast spectral clustering algorithm, whose implementation is based on efficient kernel density estimation.
\item Section~\ref{sec:user_guide}
 provides a user guide to the essential features of STAG which allow a user to apply  kernel density estimation and fast spectral clustering.
\item  Section~\ref{sec:examples} includes experiments and demonstrations of the functionality of STAG~\stagversion. 
\item   Section~\ref{sec:technical}  discusses several technical details; these include our choice of implemented algorithms, the default setup of parameters, and other technical choices. We leave these details to
the final section, as it’s not necessary for the reader to understand this when using STAG.
\end{itemize}

\subsection{Implemented Algorithms}
STAG~\stagversion\ provides an implementation of the following algorithms:

\paragraph{Euclidean Locality Sensitive Hashing.}
Given a dataset $X$ consisting of $n$ data points $x_1, \ldots, x_n \in \R^d$, and a query point $y \in \R^d$, the nearest neighbour search problem is to recover the points in $X$ close to $y$.
This forms the basis of many data analysis algorithms, and many algorithms have been developed to find approximate solutions.
Locality Sensitive Hashing~(LSH) is a tool that has been used as a building block in many such approximate nearest neighbour algorithms~\cite{andoni2006nearoptimal, andoni2015optimal}, as well as other applications~\cite{brinza2010rapid, charikar_kernel_2020, koga2007fast}.
STAG~\stagversion\ provides an efficient implementation of the Euclidean LSH scheme described by Datar et al.~\cite{lsh}.

\paragraph{Kernel Density Estimation.}
Kernel density estimation (KDE) is an important problem with many applications across machine learning and statistics.
Given $n$ data points $x_1, \ldots, x_n \in \R^d$ and a query point $q \in \R^d$, the goal of kernel density estimation is to approximate the value of 
\[
    \frac{1}{n} \sum_{i = 1}^n k(q, x_i),
\]
where $k: \R^d \times \R^d \rightarrow \R$ is a \emph{kernel function}.
The kernel density estimate can be seen as an estimate of the underlying probability density function from which the data is drawn and has been used for many applications in data science~\cite{backurs2021faster, genovese2014nonparametric, macgregor2024fast, pennington2014glove}.
STAG~\stagversion\ provides an implementation of the algorithm for kernel density estimation described by Charikar et al.\ \cite{charikar_kernel_2020}, which is based on the Euclidean locality sensitive hashing primitive.

\paragraph{Fast Spectral Clustering.}
 Kernel density estimation algorithms have been applied for fast constructions of similarity graphs used   for spectral clustering~\cite{macgregor2024fast}.
STAG~\stagversion\ provides a fast implementation of this algorithm for constructing similarity graphs and the spectral clustering algorithm.
This implementation allows spectral clustering to be applied to very large datasets in which the problem was previously intractable.

\section{Locality Sensitive Hashing} \label{sec:lsh}
Given  data points $x_1, \ldots, x_n$ in a metric space $M$ with  distance function $d: M \times M \rightarrow \R_{\geq 0}$, 
the goal of locality sensitive hashing is to preprocess the data in the way such that, given a new query point $y \in M$, an algorithm is  able to quickly recover the data points close to $y$.
Formally, a family $\mathcal{F}$ of hash functions $h: M \rightarrow S$ is \emph{locality sensitive} if
there are values $r \in \R$, $c > 1$, and $p_1 > p_2$,  such that it holds
for $h$ drawn  uniformly at random from $\mathcal{F}$ that
\[
    \p[h(u) = h(v)] \geq p_1
\]
when $d(u, v) \leq r$, and
\[
    \p[h(u) = h(v)] \leq p_2
\]
when $d(u, v) \geq c \cdot r$.
That is, the collision probability of close points is higher than that of far points.
Several techniques for locality sensitive hashing have been proposed.
Indyk and Motwani~\cite{indyk1998approximate} introduce locality sensitive hashing and propose a hashing scheme based on bit-sampling of the data vectors.
Broder~\cite{broder1997resemblance} defines the MinHash algorithm for similarity search on documents, which can be seen as a locality sensitive hashing scheme where the data are subsets of some discrete universe, such as in the bag-of-words model of documents.
Charikar~\cite{charikar2002similarity}  presents a technique based on splitting Euclidean space with random hyperplanes  and gives  a locality sensitive hashing scheme for the Euclidean space with the angular distance.
In  the STAG library and this report, we focus on the locality sensitive hashing scheme described by Datar et al.~\cite{lsh} for the Euclidean metric space.  

\subsection{Random Projection}
\begin{figure}[b]
    \centering
    \begin{subfigure}[b]{0.4\textwidth}
        \begin{center}\includegraphics[width=0.7\textwidth]{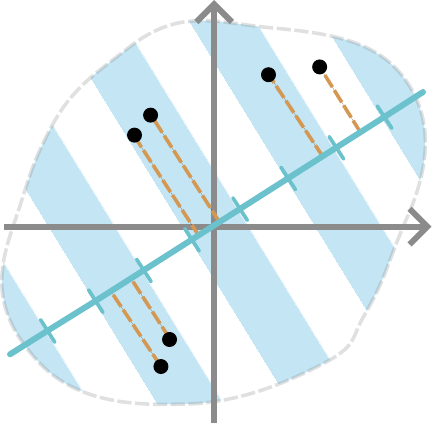}
        \caption{Projection onto a random vector.}
        \label{fig:oneproj}
        \end{center}
    \end{subfigure}
    \hfill
    \begin{subfigure}[b]{0.4\textwidth}
        \begin{center}\includegraphics[width=0.7\textwidth]{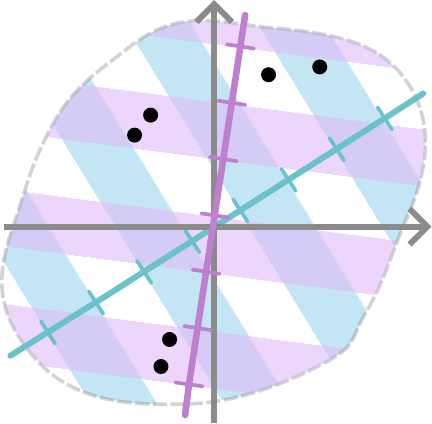}
        \caption{Projection onto two random vectors.}
        \label{fig:twoproj}
        \end{center}
    \end{subfigure}
    \caption{Demonstration of the basic unit of Euclidean LSH: projection onto a random vector. (a) Projecting onto a random vector, with discretisation into hash buckets. 
    (b) Projecting onto multiple random vectors further divides the space. Closer points are more likely to fall into the same hash bucket.}
    \label{fig:lshprojections}
\end{figure}

In this section we describe the basic hash functions used to build the Euclidean locality sensitive hashing scheme.
For some vector $a \in \R^d$ and a scalar $b \in [0, 4]$, let $h_{a, b}: \R^d \rightarrow \Z$ be the function
\[
    h_{a, b}(x) = \left\lfloor \frac{\langle x, a \rangle + b} {4} \right\rfloor.
\]
This corresponds to a projection of the vector $x \in \R^d$ onto $a$, followed by a discretisation into distinct buckets with the floor function.
The value of $b$ determines the offset of the bucket boundaries.
Figure~\ref{fig:oneproj} illustrates this function: the data is projected onto a discretised vector in order to determine the hash bucket for each point.
By selecting $a \in \R^d$ from a standard $n$ dimensional normal distribution $\mathcal{N}(0, I)$, and $b$ from the uniform distribution $U(0, 4)$, we obtain a   family $\mathcal{F}$ of hash functions.

Given two points $x_1, x_2 \in \R^d$ with $c \triangleq \| x_1 - x_2 \|$, the probability of $h(x_1) = h(x_2)$ for $h$ drawn uniformly at random from the family $\mathcal{F}$ is shown to be
\[
    p(c) \triangleq \p\left[ h(x_1) = h(x_2) \, | \, \| x_1 - x_2 \| = c \right] = \int_0^4 \frac{1}{c} f\left(\frac{t}{c}\right)\left(1-\frac{t}{4}\right)\mathrm{d}t,
\]
where $f(\cdot)$ is the probability density function of the absolute value of the normal distribution.
Solving the integral gives that
\[
    p(c)=-\frac{1}{2\sqrt{2\pi}}\left( c\cdot  \mathrm{e}^{-\frac{8}{c^2}} \right) \left( \mathrm{e}^{\frac{8}{c^2}} -1 \right) + \mathrm{erf}\left(\frac{2\sqrt{2}}{c}\right),
\]
where $\mathrm{erf}(\cdot)$ is the error function.
This function is shown in Figure~\ref{fig:oneprob}.
The STAG library provides an implementation of this LSH function.
\begin{lstlisting}[language=C++]
  #include <stag/lsh.h>
  ...
    // Create an LSH function drawn at random from the hash family F.
    StagInt dimension = 10;
    stag::LSHFunction func(dimension);

    // Apply the function to some data point.
    stag::DataPoint p = stag::DataPoint(dimension, &data);
    StagInt bucket = func.apply(p);

    // Compute the collision probability of two points at distance c.
    StagReal c = 1.5;
    StagReal p_c = stag::LSHFunction::collision_probability(c);
  ...
\end{lstlisting}

\subsection{Boosting the Probability} \label{sec:boosting}
\begin{figure}
    \centering
    \begin{subfigure}[b]{0.4\textwidth}
        \centering\includegraphics[width=0.9\textwidth]{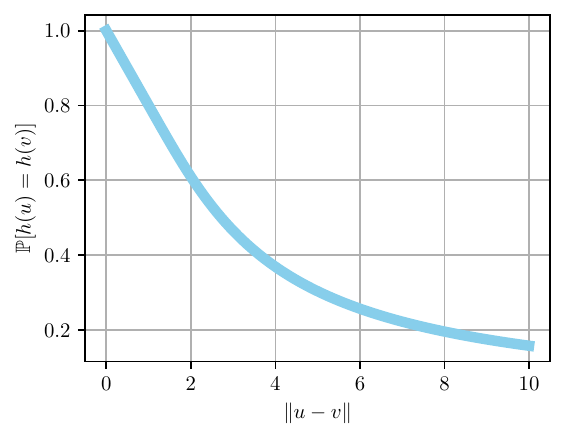}
        \caption{Collision probability $p(c)$}
        \label{fig:oneprob}
    \end{subfigure}
    \hfill
    \begin{subfigure}[b]{0.4\textwidth}
        \centering
        \includegraphics[width=0.9\textwidth]{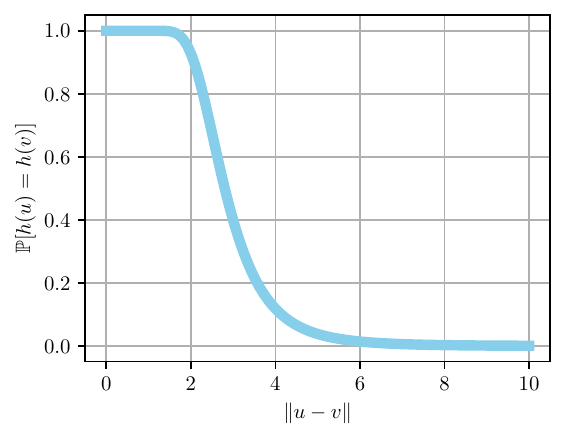}
        \caption{Collision probability for $K = 6$ and $L = 50$}
        \label{fig:multiprob}
    \end{subfigure}
    \caption{The collision probability of  two points   under a hash function drawn uniformly at random from the hash family $\mathcal{F}$. 
    Figure~(a) shows the collision probability with respect to the distance between the two input vectors $u$ and $v$; 
    Figure~(b)  shows the value of   $1 - \left(1 - p(c)^K\right)^L$, which is the collision probability when applying $K \cdot L$ independent hash functions.
    }
    \label{fig:collision_prob}
\end{figure}

By applying multiple hash functions drawn from the family $\mathcal{F}$, we can achieve different collision probabilities.
Firstly, by applying $K$ hash functions independently drawn from $\mathcal{F}$ and taking the intersections of the hash buckets, we have that points collide only if they collide under all $K$ projections.
Thus, the collision probability becomes
\[
    p(c)^K.
\]
Then, by repeating the process $L$ independent times, the probability of collision in at least one of the $L$ iterations is
\[
    1 - \left(1 - p(c)^K \right)^L.
\]
By carefully selecting $K$ and $L$, we can control the shape of the collision probability function.
Figure~\ref{fig:multiprob} shows the collision probability when $K = 6$ and $L = 50$.
The \texttt{E2LSH} class in STAG implements a Euclidean locality sensitive hash table for specified values of $K$ and $L$.

\begin{lstlisting}[language=C++]
  #include <stag/lsh.h>
  ...
    // Load a dataset from file.
    DenseMat data_mat = stag::load_matrix(filename);
    std::vector<stag::DataPoint> data = stag::matrix_to_datapoints(data_mat);

    // Create a Euclidean locality-sensitive hash table.
    StagInt K = 6;
    StagInt L = 50;
    stag::E2LSH hash_table(K, L, data);

    // Find the data points in the same bucket as some query point.
    stag::DataPoint q(data_mat, 0);
    std::vector<stag::DataPoint> close_points = hash_table.get_near_neighbors(q);
    
    // Compute the collision probability of two points at distance c.
    StagReal c = 1.5;
    StagReal p_c = hash_table.collision_probability(c);
  ...
\end{lstlisting}

\section{Kernel Density Estimation} \label{sec:kde}

\begin{figure}[htb]
    \centering
    \begin{subfigure}[b]{0.33\textwidth}
        \includegraphics[width=0.9\textwidth]{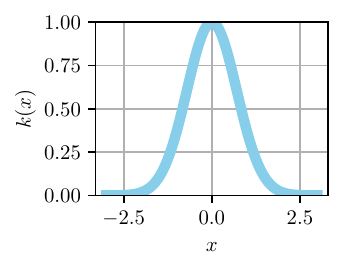}
        \caption{Gaussian kernel}
        \label{fig:gaussiankernel}
    \end{subfigure}
    \hfill
    \begin{subfigure}[b]{0.33\textwidth}
        \includegraphics[width=0.9\textwidth]{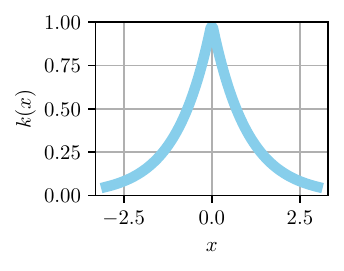}
        \caption{Exponential kernel}
        \label{fig:exponentialkernel}
    \end{subfigure}
    \hfill
    \begin{subfigure}[b]{0.33\textwidth}
        \includegraphics[width=0.9\textwidth]{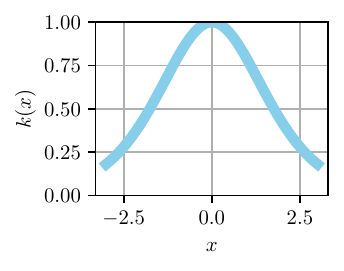}
        \caption{Logistic kernel}
        \label{fig:logistic_kernel}
    \end{subfigure}
    \caption{Three kernel functions which can be used for kernel density estimation.}
    \label{fig:kernels}
\end{figure}

\begin{figure}[thb]
    \begin{subfigure}[b]{0.3\textwidth}
        \includegraphics[width=\textwidth]{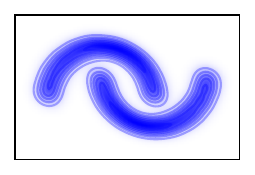}
        \caption{True data distribution}
        \label{fig:truedist}
    \end{subfigure}
    \hfill
    \begin{subfigure}[b]{0.3\textwidth}
        \includegraphics[width=\textwidth]{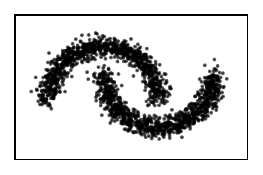}
        \caption{Data sample}
        \label{fig:datasample}
    \end{subfigure}
    \hfill
    \begin{subfigure}[b]{0.3\textwidth}
        \includegraphics[width=\textwidth]{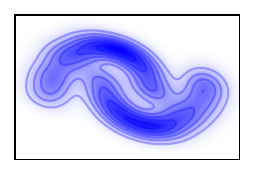}
        \caption{Kernel density estimate}
        \label{fig:kde}
    \end{subfigure}
    \caption{Kernel density estimation provides an estimate of the probability distribution from which the data is drawn.
   Figure~(a) shows the   underlying probability distribution; Figure~(b) shows the generated data points based on the probability distribution from (a); Figure~(c) shows the empirical kernel density estimate of this underlying distribution.} 
    \label{fig:kde-illustration}
\end{figure}

Given   data points $x_1, \ldots, x_n \in \R^d$, the kernel density of a query point $q \in \R^d$ is defined as 
\[
    K(q) \triangleq \frac{1}{n} \sum_{i = 1}^n k(\|q - x_i\|),
\]
where $k: \R \rightarrow \R$ is a \emph{kernel function} satisfying that 
\begin{itemize}
    \item $k(c) \in [0, 1]$,
    \item $k(c) = k(-c)$, and
    \item $k(x) \leq k(y)$ if $|x| > |y|$.
\end{itemize}
 Figure~\ref{fig:kernels} shows the plots of three typical kernel functions defined as follows:
\begin{itemize}
    \item the Gaussian kernel defined by $k(x) = \mathrm{e}^{-a x^2}$ for some bandwidth parameter $a$, 
    \item the exponential kernel  defined by $k(x) = \mathrm{e}^{-a |x|}$ for a bandwidth $a$,  and
    \item the logistic kernel defined by $k(x) = 4 / (2 + \mathrm{e}^x + \mathrm{e}^{-x})$.
\end{itemize}
 We focus on the Gaussian kernel in the remainder part of the report.
Given any dataset and an appropriate choice of kernel, the kernel density is a measure of the density of the data around a particular query point.
Figures~\ref{fig:kde-illustration} illustrates that one can see the density of the original data by computing the kernel density at every point of the plane.  This technique gives a non-parametric estimate of the probability density from which the data is drawn~\cite{epanechnikov1969nonparametric}.  

Designing efficient algorithms for computing  kernel density estimates is an active area of research in both theoretical and applied  computer science~\cite{andoni2015practical, charikar_kernel_2020, charikar2017hashing, DEANN, siminelakis2019rehashing}. 
For $n$ data points and $m$ query points in $d$ dimensions, a naive computation of  the   kernel density  requires $O(dmn)$ time.
Another natural baseline is to take a random sample of the dataset and compute an estimate of the kernel density from the sample. 
This approach gives a $(1 \pm \epsilon)$-approximation of the kernel density in time $O\left(\epsilon^{-2} \mu^{-1}  dm \log(m)\right)$, where $\mu$ is a lower bound on the kernel density of any query point~\cite{charikar_kernel_2020}.
Thus, the most interesting regime for improving the running time of kernel density estimation algorithms is when $\mu = O(1/n)$, as this is the regime in which random sampling does not improve over the naive baseline.
In low dimensions, tree-based methods~\cite{genovese2014nonparametric} and the celebrated fast Gauss transform~\cite{LJ, ifgt} perform very well, but suffer from the curse of dimensionality: their running time has   exponential dependency on $d$.
This has led to the development of methods based on locality sensitive hashing~\cite{charikar_kernel_2020, charikar2017hashing} and nearest neighbour search~\cite{DEANN}.
Among these techniques, the CKNS algorithm described by Charikar et al.\ \cite{charikar_kernel_2020} is probably the most promising one:  after the preprocessing in $\widetilde{O}\left(\epsilon^{-2}  \mu^{-0.25}   d n  \right)$ time, the algorithm gives a $(1 \pm \epsilon)$-approximation of the kernel density of $m$ query points in
$\widetilde{O}\left(\epsilon^{-2}   \mu^{-0.25}  d m  \right)$ time, where $\widetilde{O}(\cdot)$ hides poly-logarithmic factors of $n$.
Thus, when $\mu = \Theta(1/n)$, this algorithm avoids  the  exponential dependency on $d$, and  offers   significant asymptotic speed-up over the naive method and random sampling.

\subsection{The CKNS Algorithm} \label{sec:ckns}
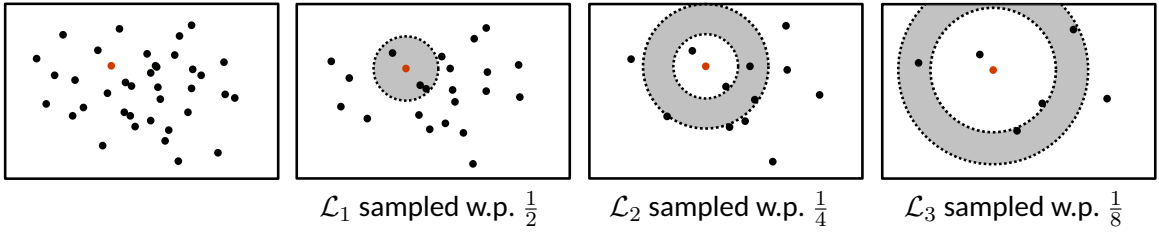
\begin{figure}[tb]
    \centering
    \hspace*{1em}
    \def\svgwidth{\textwidth}
\begingroup%
  \makeatletter%
  \providecommand\color[2][]{%
    \errmessage{(Inkscape) Color is used for the text in Inkscape, but the package 'color.sty' is not loaded}%
    \renewcommand\color[2][]{}%
  }%
  \providecommand\transparent[1]{%
    \errmessage{(Inkscape) Transparency is used (non-zero) for the text in Inkscape, but the package 'transparent.sty' is not loaded}%
    \renewcommand\transparent[1]{}%
  }%
  \providecommand\rotatebox[2]{#2}%
  \newcommand*\fsize{\dimexpr\f@size pt\relax}%
  \newcommand*\lineheight[1]{\fontsize{\fsize}{#1\fsize}\selectfont}%
  \ifx\svgwidth\undefined%
    \setlength{\unitlength}{778.35513618bp}%
    \ifx\svgscale\undefined%
      \relax%
    \else%
      \setlength{\unitlength}{\unitlength * \real{\svgscale}}%
    \fi%
  \else%
    \setlength{\unitlength}{\svgwidth}%
  \fi%
  \global\let\svgwidth\undefined%
  \global\let\svgscale\undefined%
  \makeatother%
  \begin{picture}(1,0.17512163)%
    \lineheight{1}%
    \setlength\tabcolsep{0pt}%
    \put(0,0){\includegraphics[width=\unitlength,page=1]{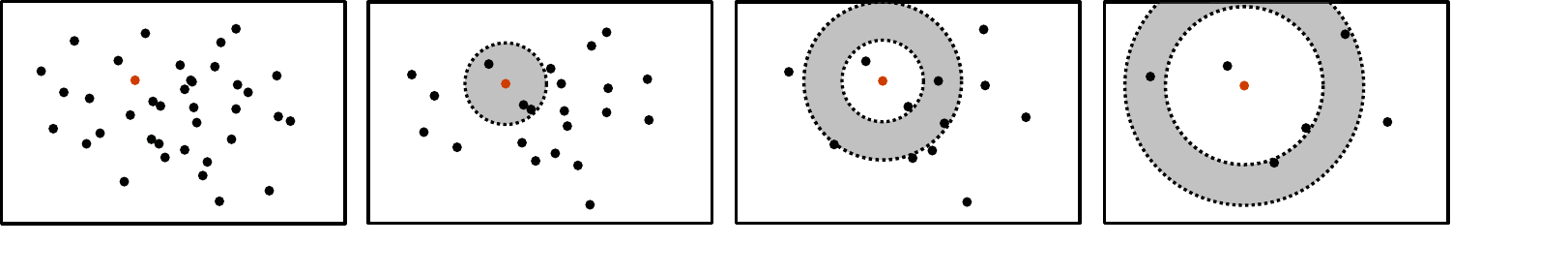}}%
    \put(0.25319974,0.00232607){\color[rgb]{0,0,0}\makebox(0,0)[lt]{\lineheight{1.25}\smash{\begin{tabular}[t]{l}$\mathcal{L}_1$ sampled w.p.\ $\frac{1}{2}$\end{tabular}}}}%
    \put(0.48565145,0.00232607){\color[rgb]{0,0,0}\makebox(0,0)[lt]{\lineheight{1.25}\smash{\begin{tabular}[t]{l}$\mathcal{L}_2$ sampled w.p.\ $\frac{1}{4}$\end{tabular}}}}%
    \put(0.72234495,0.00232607){\color[rgb]{0,0,0}\makebox(0,0)[lt]{\lineheight{1.25}\smash{\begin{tabular}[t]{l}$\mathcal{L}_3$ sampled w.p.\ $\frac{1}{8}$\end{tabular}}}}%
  \end{picture}%
\endgroup%

    \caption{The CKNS algorithm first generates several samples of the data with probabilities $ 1/2, 1/4, \ldots, 1/n $.
    Then,  for any query point $q$~(indicated as the red point shown above), the data points in $\mathcal{L}_i$ are recovered from the data sampled with probability $2^{-i}$.}
    \label{fig:ckns}
\end{figure}

In this section, we give a high-level description of the CKNS kernel density estimation algorithm,  and refer the reader to \cite{charikar_kernel_2020} for the detailed discussion and analysis of the algorithm.
The CKNS algorithm is based on the common algorithmic pattern of importance sampling.
Intuitively, when computing the kernel density of some point $q$, the data points closer to $q$ have more influence on the kernel density.
As such, we would like to select a random sample of the data points with the sampling probability depending on the distance to the query point.
To this end, for some query point $q$, we define the sets $\mathcal{L}_i \subseteq \{x_1, \ldots, x_n\}$ such that it holds for all $x_j \in \mathcal{L}_i$ that 
\[
    2^{-i} \leq k(\|q - x_j\|) \leq 2^{-i+1}.
\]
Notice that, for $i < j$, the points in $\mathcal{L}_i$
are closer to $q$ than the points in $\mathcal{L}_j$.
Hence, by sampling the points in $\mathcal{L}_i$ with probability proportional to $2^{-i}$, we can obtain an estimate of the kernel density of $q$.
To achieve this, the CKNS algorithm proceeds with the following steps, which are illustrated in Figure~\ref{fig:ckns}.
\begin{enumerate}
    \item Create $\log n$ subsampled datasets, using sampling probabilities $\left\{2^{-i}\right\}_{i=1}^{\log n}$.
    \item For each subsampled dataset, create a locality sensitive hash table to recover points in $\mathcal{L}_i$ for any query point.
\end{enumerate} 
Then, to estimate the kernel density of a given query point $q$,   we use the locality sensitive hash table to recover the points in each level $\mathcal{L}_i$ from the dataset sampled with probability  $2^{-i}$, and the kernel density estimate for $q$ is
\[
    \sum_{i = 1}^{\log n} \sum_{\substack{x \in \mathcal{L}_i \\ x \text{ sampled w.p. } 2^{-i}}} 2^i \cdot k(\|q - x\|).
\]
A direct calculation shows that   this gives an unbiased estimate for the kernel density of $q$.
The algorithm is repeated $O(\log n)$ times in order to return a more accurate estimate.

\subsection{Kernel Density 
Estimation in STAG}
STAG~\stagversion\ provides an implementation of the CKNS KDE algorithm which is intended to be both efficient and easy to use.

\begin{lstlisting}[language=C++]
  #include <stag/kde.h>
  ...
    // Load a dataset from file.
    DenseMat data_mat = stag::load_matrix(filename);
    std::vector<stag::DataPoint> data = stag::matrix_to_datapoints(data_mat);

    // Construct a KDE data structure.
    StagReal a = 0.01;
    stag::CKNSGaussianKDE kde(data, a);

    // Estimate the kernel density of some query point.
    stag::DataPoint q(data_mat, 0);
    StagReal density = kde.query(q);
  ...
\end{lstlisting}

\section{Fast Spectral Clustering} \label{sec:sc}
 
For any set of  points $X\triangleq \{x_1, \ldots, x_n\} \subset \R^d$ and parameter $k\in\mathbb{N}$, the goal of spectral clustering is to partition the $n$ points into $k$ clusters such that similar
points are grouped into the same cluster.
Spectral clustering consists of the  following three steps:
\begin{enumerate}
\item compute a similarity graph $G$ of $n$ vertices;
\item compute the bottom  $k$ eigenvectors of the Laplacian matrix of $G$, and apply these eigenvectors to embed the vertices into $\mathbb{R}^k$;
\item apply $k$-means on the embedded points, and use   $k$-means' output as the resulting clustering.
\end{enumerate}
There are several methods for constructing a similarity graph in the first step~\cite{von2007tutorial}.
One popular method to construct a fully connected similarity graph $G$ is to use a kernel function $K: \R \rightarrow \R$ in the following way:
every $x_i\in X$ corresponds to a vertex $i$ in $G$, and any pair of vertices $i$ and $j$ is connected by an edge with weight $K(\|x_i - x_j\|)$. 
Other constructions of similarity graphs include 
$k$-nearest neighbour graphs and $\epsilon$-neighbourhood graphs, and more information can be found in \cite{ngSpectralClusteringAnalysis2001, shi2000normalized, von2007tutorial}.

\subsection{The MS Algorithm\label{sec:ms}}
 
One significant downside to the fully connected similarity graph is that the adjacency matrix is dense, and constructing this graph requires $\Theta(n^2)$ time and space complexity. To overcome this,   Macgregor and Sun present an algorithm that constructs a similarity graph in $\widetilde{O}(T_{\mathrm{KDE}}(n))$ time,
where $T_{\mathrm{KDE}}(n)$ is the time taken to compute approximate kernel density estimates for every point in the data set.
The constructed graph is proven to have the same cluster structure as the fully connected one~\cite{macgregor2024fast}; we call their algorithm the MS algorithm.
For low-dimensional data~\footnote{In theory, ``low-dimensional'' means that $d = O(1)$. In practice, the fast Gauss transform can be applied when $d \leq 5$.}, the \emph{fast Gauss transform} can be applied to compute the kernel density estimates in $O(n)$ time, and Macgregor and Sun provide an implementation of their algorithm in this regime.

To give a high level description of the MS algorithm, assume that $F=(V_F, E_F, w_F)$ is the fully connected graph constructed from $X$ with the kernel function $K$, and $F$ has $k$ well-defined clusters.
It is known that, if one samples every edges $e=(i,j)\in E_F$ with probability proportional to 
\begin{equation}\label{eq:sampling_prob}
\frac{w_F(i,j)}{\deg_F(i)} + \frac{w_F(i,j)}{\deg_F(j)}, 
\end{equation}
where 
\[
    \deg_F(i) = \sum_{j = 1}^n w_F(i, j) = \sum_{j = 1}^n K(\|x_i - x_j\|)
\] is the degree of $i$ in $F$, 
then the resulting graph $G$  has $\widetilde{O}(n)$ edges and   the same cluster structure as $F$~\cite{SZ19}.
Since $\deg_F(i)$ for every   $i$ is proportional to the kernel density at $x_i$,
the MS algorithm demonstrates that, by applying a KDE algorithm as a black-box,
one can sample every edge $(i,j)$ with approximately the same probability as \eqref{eq:sampling_prob}.
Based on this technique, the MS algorithm avoids the computation of all the edge weights in $F$, and achieves the total running time of $\widetilde{O}(T_{\mathrm{KDE}})$.

\subsection{Fast Spectral Clustering in STAG}
 
STAG~\stagversion\ includes an implementation of the MS algorithm~\cite{macgregor2024fast}  based on the fast KDE algorithm in Section~\ref{sec:kde}.
The implementation is very easy to use, and requires only the data matrix and the parameter $a$ of the Gaussian kernel.

\begin{lstlisting}[language=C++]
  #include <stag/cluster.h>
  #include <stag/graph.h>
  ...
    // Load a dataset from file.
    DenseMat data_mat = stag::load_matrix(filename);
    std::vector<stag::DataPoint> data = stag::matrix_to_datapoints(data_mat);

    // Construct an approximate similarity graph.
    StagReal a = 0.01;
    stag::Graph g = stag::approximate_similarity_graph(&data, a);

    // Perform spectral clustering on the constructed graph.
    StagInt k = 2;
    std::vector<StagInt> labels = stag::spectral_cluster(&g, k);
  ...
\end{lstlisting}

\section{User's Guide} \label{sec:user_guide}
In this section, we include a detailed user's guide for the new features added to STAG in this release.
For the installation instructions and other guidance on using STAG, see our first technical report~\cite{stag1} and the online documentation\footnote{\url{https://staglibrary.io/}}.
Although the examples in this section use C++, the same functionality is available in STAG Python.
Appendix~\ref{app:ug_python} includes example code showing how to perform locality sensitive hashing, kernel density estimation, and spectral clustering with STAG Python.

\subsection{Handling Data} \label{sec:ug_data}
All of the added functionality in this release is based on data sets in $\R^d$.
STAG~\stagversion\ introduces the \lstinline{DenseMat} type alias which is used to represent a dense matrix in $\R^{n \times d}$ containing the $n$ data points.
\begin{lstlisting}[language=C++]
    typedef Eigen::Matrix<double, Eigen::Dynamic, Eigen::Dynamic, Eigen::RowMajor> DenseMat
\end{lstlisting}
STAG provides methods for reading and writing such data matrices to disk.
If the data is stored in a space or comma separated file with one
data point on each line, then the \texttt{stag::load\_matrix} method can be used to read the file.
\begin{lstlisting}[language=C++]
    #include <stag/data.h>
    ...
        std::string my_filename = "data.csv";
        DenseMat data = stag::load_matrix(my_filename);
    ...
\end{lstlisting}
Furthermore, the \texttt{stag::save\_matrix} method can be used to save a \texttt{DenseMat} to disk.
\begin{lstlisting}
    ...
        stag::save_matrix(data, my_filename);
    ...
\end{lstlisting}
Many methods in the STAG library take a single data point as an argument rather than the entire dataset.
For this purpose, STAG~\stagversion\ introduces the \texttt{stag::DataPoint} class, which is a wrapper around a pointer to the underlying data.
The most common use case of the \texttt{stag::DataPoint} class is to 
point to a certain row of a \texttt{DenseMat}.
\begin{lstlisting}[language=C++]
    #include <stag/data.h>
    ...
        DenseMat data = stag::load_matrix(my_filename)

        // Create a data point referencing the 10th row of the data matrix.
        stag::DataPoint point(data, 10);
    ...
\end{lstlisting}
A \texttt{stag::DataPoint} object can also be initialised using a C++ vector or by passing a pointer to the data directly.

\subsection{Locality Sensitive Hashing} \label{sec:ug_lsh}
The \texttt{lsh.h} module provides the Euclidean locality sensitive hashing algorithm.

\paragraph{The LSHFunction Class.}
The \texttt{stag::LSHFunction} class represents a single locality sensitive hash function based on projection onto a random vector.
The function is initialised with a single argument specifying the dimension of the data.
\begin{lstlisting}[language=C++]
    #include <stag/lsh.h>
    ...
      // Create an LSH function for 10-dimensional data.
      stag::LSHFunction f(10);
    ...
\end{lstlisting}
The \texttt{apply} method is used to compute the hashed value of a data point.
\begin{lstlisting}[language=C++]
    ...
        // Create a data point referencing the 10th row of the data matrix.
        stag::DataPoint p(data, 10);

        // Compute the hash value of p.
        StagInt h = f.apply(p);
    ...
\end{lstlisting}

\paragraph{The E2LSH Class.}
The \texttt{stag::E2LSH} class represents a hash table constructed by combining several hash functions.
As described in Section~\ref{sec:boosting}, by combining hash functions with the parameters $K$ and $L$, we can control the collision probability of points hashed into the hash table.
The \texttt{stag::E2LSH} class takes $K$ and $L$ as arguments, along with a vector of \texttt{stag:DataPoint} objects to be stored in the table.
\begin{lstlisting}[language=C++]
    #include <stag/lsh.h>
    #include <stag/data.h>
    ...
        // Load the data points.
        DenseMat data = stag::load_matrix(my_filename)
        std::vector<stag::DataPoint> points = stag::matrix_to_datapoints(&data);

        // Create the E2LSH hash table.
        StagInt K = 6;
        StagInt L = 50;
        stag::E2LSH table(K, L, points);
    ... 
\end{lstlisting}
Once the hash table is constructed, the approximate near neighbours of a query point can be returned with the \texttt{get\_near\_neighbours} method.
\begin{lstlisting}[language=C++]
    ...
        // Use the 10th row of the data matrix as the query point.
        stag::DataPoint q(data, 10);

        // Return the data points with the same hash value as the query.
        std::vector<stag::DataPoint> close_points = table.get_near_neighbours(q);
    ...
\end{lstlisting}

\subsection{Kernel Density Estimation} \label{sec:ug_kde}
The \texttt{kde.h} module provides methods for \textsf{KDE} with   Gaussian kernel, and  \texttt{stag::gaussian\_kernel}   computes the Gaussian kernel value between two data points.
\begin{lstlisting}[language=C++]
    #include <stag/kde.h>
    #include <stag/data.h>
    ...
        // Load some data.
        DenseMat data = stag::load_matrix(my_filename);
        stag::DataPoint p1(data, 1);
        stag::DataPoint p2(data, 2);

        // Compute the Gaussian kernel similarity between the data points.
        StagReal a = 1;
        StagReal val = stag::gaussian_kernel(a, p1, p2);
    ...
\end{lstlisting}

\paragraph{The ExactGaussianKDE Class.}
The \texttt{stag::ExactGaussianKDE} class gives a method to compute the Gaussian kernel density exactly.
The data structure is initialised by passing the parameter $a$ of the Gaussian kernel and a \texttt{DenseMat} with the dataset.
\begin{lstlisting}[language=C++]
    #include <stag/kde.h>
    #include <stag/data.h>
    ...
        // Load the dataset.
        DenseMat data = stag::load_matrix(my_filename);

        // Create the kernel density data structure.
        StagReal a = 1;
        stag::ExactGaussianKDE kde(&data, a);
    ...
\end{lstlisting}
The kernel density for a query point can then be computed with the \texttt{query} method.
\begin{lstlisting}[language=C++]
    ...
        stag::DataPoint q(data, 10);
        StagReal kernel_density = kde.query(q);
    ...
\end{lstlisting}
To query multiple points simultaneously, we can instead pass a \texttt{DenseMat} to the query method to obtain the kernel density for every row of the matrix.
\begin{lstlisting}[language=C++]
    ...
        std::vector<StagReal> kernel_densities = kde.query(&data);
    ...
\end{lstlisting}
The time complexity of this method is $O(d m n )$ where $m$ is the number of query points, $n$ is the number of data points, and $d$ is the dimensionality of the data.

\paragraph{The CKNSGaussianKDE Class.}
The \texttt{stag::CKNSGaussianKDE}   class provides kernel density estimates with the CKNS algorithm described in Section~\ref{sec:ckns}.
There're 3   methods to initialise a \texttt{stag::CKNSGaussianKDE} object.
The simplest method is to initialise with only a data matrix and the parameter $a$ of the Gaussian kernel.
\begin{lstlisting}[language=C++]
    #include <stag/kde.h>
    #include <stag/data.h>
    ...
        // Load the dataset.
        DenseMat data = stag::load_matrix(my_filename);

        // Create the KDE data structure.
        StagReal a = 1;
        stag::CKNSGaussianKDE kde(&data, a);
    ...
\end{lstlisting}
This method will choose sensible default parameters for the CKNS algorithm and give good kernel density estimates.
For more control over the trade-off between time complexity and accuracy, the \texttt{eps} and \texttt{min\_mu} arguments can be used.
\begin{itemize}
    \item The \texttt{eps} argument corresponds to the $\epsilon$ error parameter of the CKNS KDE data structure, which is designed to return estimates within a $(1 \pm \epsilon)$-factor of the exact kernel density.
    Recall that the initialisation time complexity of the data structure is $\widetilde{O}(\varepsilon^{-2} n^{1.25})$ and the query time complexity is $\widetilde{O}(\epsilon^{-2} n^{0.25})$. The default value is $0.5$.
    \item The \texttt{min\_mu} argument is an estimated minimum kernel density value of any query point.
    A smaller number will give longer pre-processing and query time complexity.
    If a query point has a true kernel density smaller than this value, then the data structure may return an inaccurate estimate.
    The default value is $1 / n$.
\end{itemize}
\begin{lstlisting}[language=C++]
    ...
        // Create the KDE data structure.
        StagReal eps = 0.5;
        StagReal a = 1;
        StagReal min_mu = 0.0001;
        stag::CKNSGaussianKDE kde(&data, a, eps, min_mu);
    ...
\end{lstlisting}
For those familiar with the details of the CKNS algorithm, the final constructor allows fine-grained control over the constants used by the data structure to control the accuracy and variance of the estimator.
\begin{itemize}
    \item The \texttt{K1} parameter specifies the number of independent copies of the data structure to create in parallel. This controls the variance of the estimates returned by the \texttt{query} method.
    It is usually set to $O(\epsilon^{-2} \log(n))$.
    \item The \texttt{K2\_constant} parameter controls the number of hash functions used in the E2LSH hash tables within the data structure.
    A higher value will reduce the variance of the estimates at the cost of higher memory and time complexity.
    It is usually set to $O(\log(n))$.
    \item The CKNS algorithm samples the data points with different sampling probabilities.
    Setting   \texttt{p\_offset}   to $k \geq 0$ will further subsample the data by a factor of $1 / 2^k$.
    This will speed up the algorithm at the cost of some accuracy.
    It is usually set to $0$.
\end{itemize}
\begin{lstlisting}[language=C++]
    ...
        // Create the KDE data structure.
        StagReal a = 1;
        StagReal min_mu = 0.0001;
        StagInt k1 = 10;
        StagReal k2_constant = 5;
        StagInt p_offset = 1;
        stag::CKNSGaussianKDE kde(&data, a, min_mu, k1, k2_constant, p_offset);
    ...
\end{lstlisting}
Once the CKNS data structure has been initialised, the kernel density estimate for any query point can be computed with the \texttt{query} method.
\begin{lstlisting}[language=C++]
    ...
        stag::DataPoint q(data, 10);
        StagReal kd_estimate = kde.query(q);
    ...
\end{lstlisting}
To query multiple data points, it is more efficient to pass a \texttt{DenseMat} containing the query points as rows.
\begin{lstlisting}[language=C++]
    ...
        std::vector<StagReal> kd_estimates = kde.query(&data);
    ...
\end{lstlisting}

\subsection{Spectral Clustering} \label{sec:ug_sc}
STAG~\stagversion\ introduces the new \texttt{stag::approximate\_similarity\_graph} method for constructing
a similarity graph from data.
The method uses the MS algorithm with the Gaussian kernel and the CKNS kernel density estimation structure.
It requires only two arguments: the \texttt{DenseMat} containing the data, and the parameter $a$ of the Gaussian kernel.
\begin{lstlisting}[language=C++]
    #include <stag/cluster.h>
    #include <stag/data.h>
    #include <stag/graph.h>
    ...
        // Load the dataset.
        DenseMat data = stag::load_matrix(my_filename);

        // Create the approximate similarity graph.
        StagReal a = 1;
        stag::Graph g = stag::approximate_similarity_graph(&data, a);
    ...
\end{lstlisting}
When the dataset has a clear cluster structure, this method is guaranteed to preserve the structure of the fully connected similarity graph, which can be constructed with the \texttt{stag::similarity\_graph} method.
Once the similarity graph is constructed, we can find the clusters using the \texttt{stag::spectral\_cluster} method.
\begin{lstlisting}[language=C++]
    ...
        // Find 10 clusters in the graph.
        StagInt k = 10;
        std::vector<StagInt> labels = stag::spectral_cluster(&g, k);
    ...
\end{lstlisting}

\newcommand{\aloi}{\texttt{aloi}}
\newcommand{\glove}{\texttt{glove}}
\newcommand{\shuttle}{\texttt{shuttle}}
\newcommand{\msd}{\texttt{msd}}
\newcommand{\mnist}{\texttt{mnist}}
\newcommand{\covtype}{\texttt{covtype}}

\section{Showcase Studies} \label{sec:examples}
In this section, we demonstrate the performance of the kernel density estimation and spectral clustering algorithms available with STAG through experiments on real-world and synthetic datasets.
The kernel density estimation experiments are performed on a compute server with 64 AMD EPYC 7302 16-Core Processors and 500 Gb of RAM and the spectral clustering experiments are performed on a laptop with an 11th Gen Intel(R) Core(TM) i7-11800H @ 2.30GHz processor and 32 GB RAM.
The code used to produce all experimental results is available at  
\begin{equation*}
\mbox{\url{https://github.com/staglibrary/kde-experiments}.}
\end{equation*}

\subsection{Kernel Density Estimation}
In this section, we compare the STAG implementation of the CKNS KDE algorithm~\cite{charikar_kernel_2020} with the following alternative methods for kernel density estimation.

\begin{itemize}
    \item \textbf{Naive}: compute the exact kernel density by measuring the kernel similarity with every data point.
    \item \textbf{Random sampling (rs)}: estimate the kernel density by sampling a uniformly random subset of the data.
    \item \textbf{DEANN}~\cite{DEANN}: importance sampling using an approximate nearest neighbour algorithm as a black box.
    \item \textbf{Sklearn}~\cite{scikit-learn}: kernel density estimation with the ball-tree algorithm implemented in the sklearn machine learning library.
\end{itemize}

We use the implementation of the naive, random sampling, and DEANN algorithms provided by Karppa et al.\ \cite{DEANN}.
They provide two implementations of the random sampling, and DEANN algorithms: a basic implementation, and a ``permuted'' implementation with improved memory efficiency which we refer to as ``random sampling permuted'' (rsp), and DEANNP.
We evaluate the algorithms on the following datasets, whose sizes are reported in Table~\ref{tab:dataset-sizes}:

\begin{wraptable}[5]{r}{5cm}
    \centering \caption{Values of $n$ and $d$ for the tested datasets}
    \begin{tabular}{rrr}
        \toprule
        dataset & $n$ & $d$ \\
        \midrule
        aloi & 108,000 & 128 \\
        covtype & 581,012 & 54 \\
        glove & 1,193,514 & 100 \\
        mnist & 70,000 & 728 \\
        msd & 515,345 & 90 \\
        shuttle & 58,000 & 9 \\
        \bottomrule
    \end{tabular}
    \label{tab:dataset-sizes}
\end{wraptable}
\ 
\vspace{-\baselineskip}
\begin{itemize}
    \item \textbf{aloi}~\cite{geusebroek2005amsterdam}: colour images of objects under a variety of lighting conditions.
    \item \textbf{covtype}~\cite{covtype_dataset}: numerical and categorical data about land use.
    \item \textbf{glove}~\cite{pennington2014glove}: word embedding vectors.
    \item \textbf{mnist}~\cite{lecun_mnist_1998}: greyscale images of handwritten digits.
    \item \textbf{msd}~\cite{bertin2011million}: numerical and categorical metadata of songs.
    \item \textbf{shuttle}~\cite{shuttle_148}: numerical data from sensors on the space shuttle.
\end{itemize}
We follow the experimental setup established by Karppa et al.\ \cite{DEANN}.
For each dataset, we select 10,000 points uniformly at random to form the query set and perform a grid search over the parameter space of each algorithm to compute the kernel density estimates of the query points.
For each set of parameters, we measure the average relative error defined by 
\[
    \frac{1}{m} \sum_{i = 1}^m \left| \frac{\mu_i - \mu_i^*} {\mu_i^*} \right|,
\]
where $m$ is the size of the query set, $\mu_i$ is the kernel density estimate of the $i$th query point, and $\mu_i^*$ is the true kernel density.
In Table~\ref{tab:kde_times} we report the lowest query time in milliseconds of all parameter configurations which achieve a relative error below $0.1$. Figure~\ref{fig:times} further shows how the optimal running time varies for different relative errors.
From these results, we observe that the STAG implementation is faster than our tested algorithms for all datasets and relative errors.

\newcommand{\timestablecaption}{Minimum per-query time in milliseconds achieving a relative error less than $0.1$}
\begin{table}[htb]
    \caption{\timestablecaption}
    \label{tab:kde_times}
    \centering
    \begin{tabular}{lcccccccc}
    \toprule
    & \multicolumn{7}{c}{Algorithm} \\ 
    \cmidrule(l){2-8} 
    dataset & stag & DEANN & DEANNP & naive & rs & rsp & sklearn\\ 
    \midrule
    aloi &  0.017 &  0.306 &  0.088 &  0.448 &  0.341 &  0.077 &  26.206 \\
    covtype &  0.019 &  0.934 &  0.325 &  3.124 &  0.958 &  0.141 &  158.877 \\
    glove &  0.001 &  0.030 &  0.006 &  7.019 &  0.027 &  0.013 &  304.694 \\
    mnist &  0.014 &  0.259 &  0.082 &  0.189 &  0.117 &  0.081 &  35.009 \\
    msd &  0.011 &  0.610 &  0.267 &  1.758 &  3.355 &  0.743 &  124.321 \\
    shuttle &  0.022 &  0.126 &  0.081 &  0.144 &  0.445 &  0.034 &  1.919 \\
    \bottomrule
  \end{tabular}
\end{table}

\newcommand{\timesfigcaption}{Comparison of running time against relative error.}
\begin{figure}[htb!]
    \centering
    \begin{subfigure}[b]{0.47\textwidth}
        \begin{center}\includegraphics[width=\textwidth]{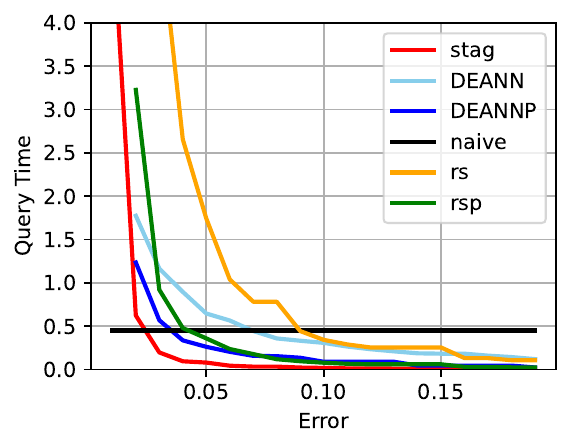}
        \caption{\aloi}
        \label{fig:aloi01}
        \end{center}
    \end{subfigure}
\hfill
    \begin{subfigure}[b]{0.49\textwidth}
        \begin{center}\includegraphics[width=\textwidth]{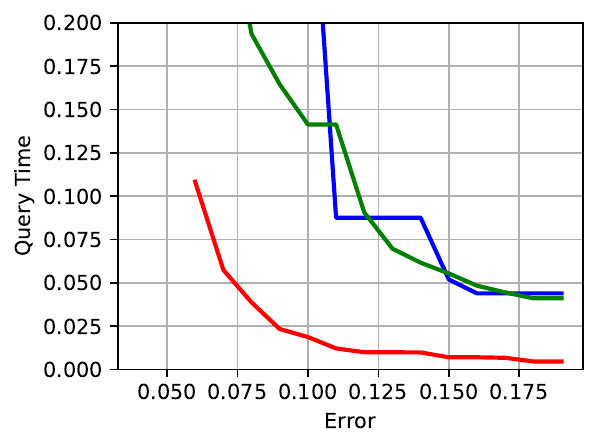}
        \caption{\covtype}
        \label{fig:covtype01}
        \end{center}
    \end{subfigure}
\hfill
    \begin{subfigure}[b]{0.49\textwidth}
        \begin{center}\includegraphics[width=\textwidth]{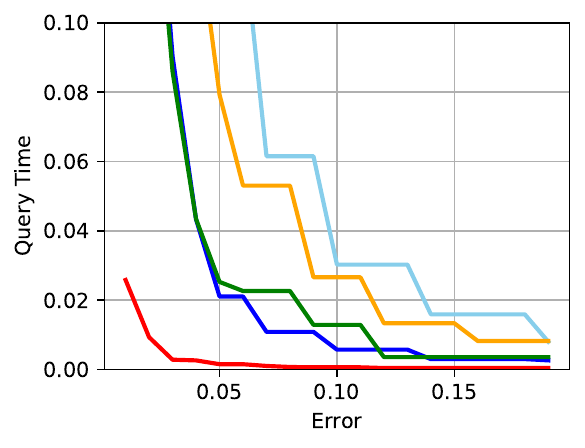}
        \caption{\glove}
        \label{fig:glove01}
        \end{center}
    \end{subfigure}
\hfill
    \begin{subfigure}[b]{0.49\textwidth}
        \begin{center}\includegraphics[width=\textwidth]{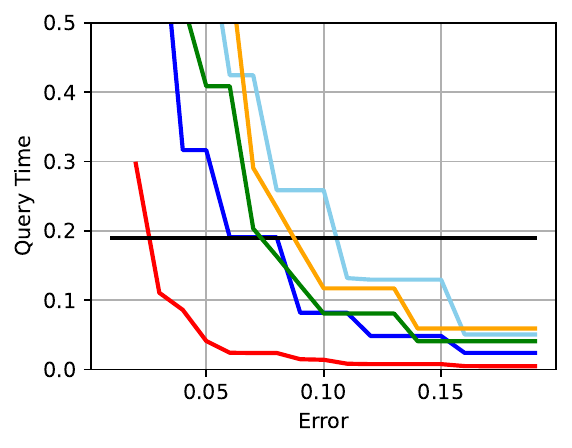}
        \caption{\mnist}
        \label{fig:mnist01}
        \end{center}
    \end{subfigure}
\hfill
    \begin{subfigure}[b]{0.49\textwidth}
        \begin{center}\includegraphics[width=\textwidth]{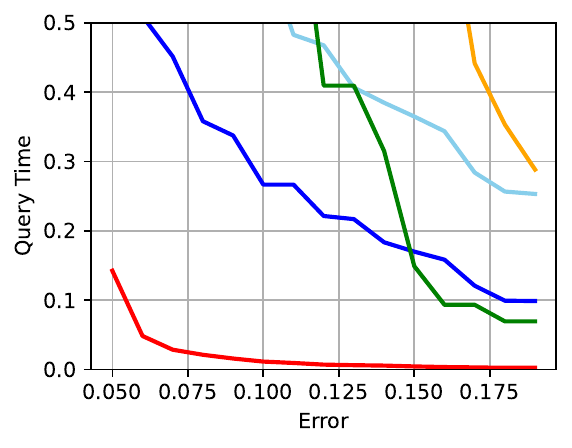}
        \caption{\msd}
        \label{fig:msd01}
        \end{center}
    \end{subfigure}
\hfill
    \begin{subfigure}[b]{0.49\textwidth}
        \begin{center}\includegraphics[width=\textwidth]{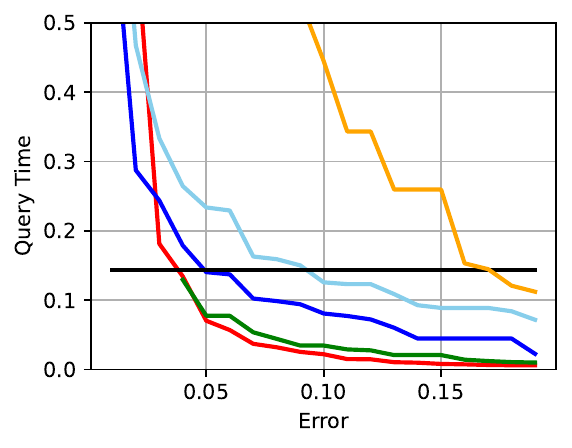}
        \caption{\shuttle}
        \label{fig:shuttle01}
        \end{center}
    \end{subfigure}
    \caption{\timesfigcaption}
    \label{fig:times}
\end{figure}


\subsection{Spectral Clustering}
In this section, we compare several spectral clustering implementations based on a variety of similarity graph constructions.

\begin{itemize}
    \item \textbf{Sklearn FC}: the fully connected similarity graph constructed with the sklearn Python library~\cite{scikit-learn}.
    \item \textbf{Sklearn kNN}: the $k$-nearest neighbour similarity graph constructed with the sklearn Python library~\cite{scikit-learn}.
    \item \textbf{FAISS HNSW}: an approximate $k$-nearest neighbour similarity graph constructed by the hierarchical navigable small worlds algorithm~\cite{malkov2018efficient} from the FAISS library~\cite{douze2024faiss}.
    \item \textbf{FAISS IVF}: an approximate $k$-nearest neighbours similarity graph constructing using the inverted file algorithm from the FAISS library~\cite{douze2024faiss}.
    \item \textbf{MS}: the similarity graph construction algorithm  described in Section~\ref{sec:ms}
    based on kernel density estimation with the fast Gauss transform~\cite{ifgt}.
    \item \textbf{STAG}: the similarity graph algorithm provided by STAG, which is based on the MS algorithm and uses the CKNS algorithm~\cite{charikar_kernel_2020} for kernel density estimation.
\end{itemize}

We evaluate the algorithms on two synthetic datasets:
\begin{itemize}
    \item \textbf{moons}: the two-moons dataset from the sklearn library. The data has two dimensions and consists of  two clusters which are not linearly separable.
    \item \textbf{blobs}: data is generated from a mixture of Gaussian distributions using the sklearn library. We   fix the number of clusters to be 10, and   vary the number of dimensions and the number of data points.
\end{itemize}

\begin{figure}[htbp]
    \centering
    \begin{subfigure}[b]{0.49\textwidth}
        \begin{center}\includegraphics[width=\textwidth]{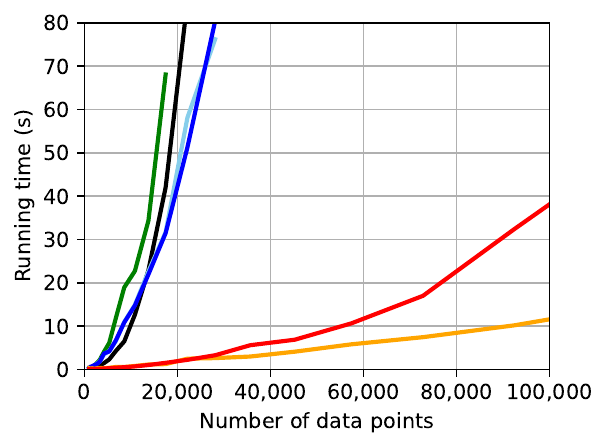}
        \caption{Comparison on the two-dimensional moons dataset.}
        \label{fig:moons}
        \end{center}
    \end{subfigure}
\hfill
    \begin{subfigure}[b]{0.49\textwidth}
        \begin{center}\includegraphics[width=0.95\textwidth]{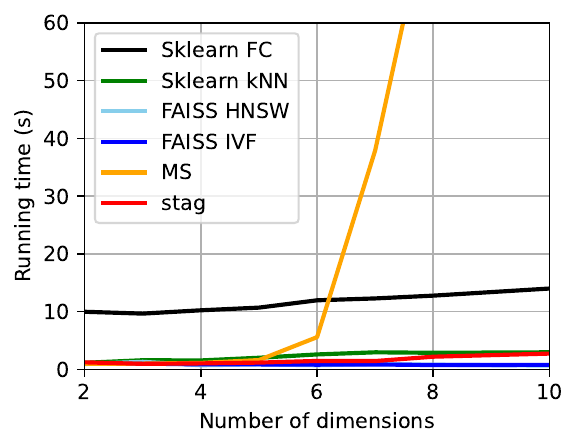}
        \caption{Comparison on the blobs dataset with 10,000 points.}
        \label{fig:blobs_dim}
        \end{center}
    \end{subfigure}
    \caption{Comparison of different clustering algorithms' running time for low-dimensional data.  For every input instance, all the algorithms perfectly return the ground truth clustering.}
    \label{fig:low_dim}
\end{figure}

The running time of the fast Gauss transform has   exponential dependency on the dimension, and as such we expect the MS algorithm~\cite{macgregor2024fast} to perform well only in low dimensions.
To compare the algorithms' performance in the low dimensional setting, we first follow the experimental setup of Macgregor and Sun~\cite{macgregor2024fast} on the two-dimensional two moons dataset.
We vary the number of data points and report the running time of each algorithm in Figure~\ref{fig:moons}.
Secondly, we use the blobs dataset to study how the performance of each algorithm changes as the  dimension of data points increases.
We fix the number of data points to be 10,000 and vary the dimensionality of the data.
The running time of each algorithm is shown in Figure~\ref{fig:blobs_dim}.
From these results, we can see that in the two-dimensional case, the MS algorithm outperforms STAG.
On the other side, as the dimension of data points increases, the MS algorithm suffers from   exponential increase of its   running time, while the running time of the other algorithms is not significantly affected.

\begin{figure}[htb]
    \centering
    \begin{subfigure}[b]{0.49\textwidth}
        \begin{center}\includegraphics[width=\textwidth]{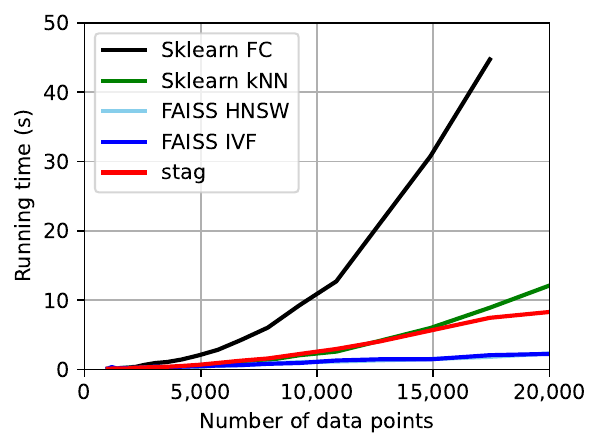}
        \caption{Experimental results on smaller values of $n$}
        \label{fig:blobssmall}
        \end{center}
    \end{subfigure}
\hfill
    \begin{subfigure}[b]{0.49\textwidth}
        \begin{center}\includegraphics[width=\textwidth]{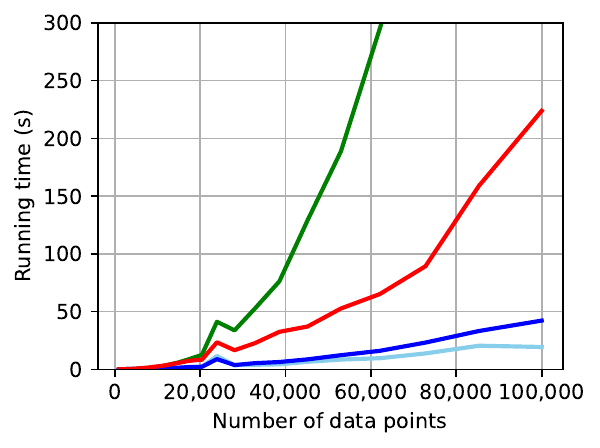}
        \caption{Experimental results on larger values of  $n$}
        \label{fig:blobsn}
        \end{center}
    \end{subfigure}
    \caption{Running time comparison of different clustering algorithms on 100-dimensional data from the blobs dataset. For every input instance, all the algorithms perfectly return the ground truth clustering. }
    \label{fig:blobs_n}
\end{figure}

Finally, we evaluate the algorithms, excluding MS, on 100-dimensional data from the blobs dataset and report the running time in Figure~\ref{fig:blobs_n}.
 We can see that the STAG algorithm has good performance for high-dimensional data although the FAISS algorithms run 
 faster when the number of data points is large.
However, it is worth noting that the HNSW and IVF algorithms are based on heuristics for approximate nearest-neighbour search without formal theoretical guarantees.
Moreover, the STAG algorithm is designed to approximate the \emph{fully-connected} similarity graph rather than the $k$-nearest neighbour graph.
Hence, among the compared algorithms, the STAG algorithm is the only one with good performance on both low and high dimensional datasets, as well as  offering theoretical guarantee on it performance.

\section{Technical Considerations} \label{sec:technical}
In this section we discuss some of the technical choices made in the design and implementation of STAG.

\paragraph{Implementation of Euclidean LSH.}
The Euclidean LSH algorithm is implemented from scratch in STAG, although the existing C implementation by Andoni~\cite{E2LSHmanual} is consulted for reference, and we are grateful to Andoni and Indyk for making this implementation available.
We choose to re-implement the algorithm in order to take advantage of some modern features of C++ including the \texttt{std::vector} data structure.
Re-implementing Euclidean LSH also makes us easier to expose it in the STAG Python library.

\paragraph{Choice of $b$ in Euclidean LSH implementation.}
In the STAG implementation of the Euclidean LSH function, we select the parameter $b$ uniformly at random from the interval $[0, 4]$.
The choice of $4$ is   arbitrary, and choosing another constant will also work with   a different corresponding collision probability function.
On the use of  the interval $[0, 4]$, we follow the recommendation of Andoni~\cite{E2LSHmanual}.

\paragraph{Choice of implemented KDE algorithm.}
In recent years, several kernel density algorithms have been proposed~\cite{charikar2017hashing, charikar_kernel_2020, DEANN,  siminelakis2019rehashing}.
Of these, we choose to implement the CKNS algorithm~\cite{charikar_kernel_2020} because it is a theoretically grounded algorithm which is also relatively simple. 
Furthermore, prior to the release of STAG~\stagversion\ it lacked a publicly available implementation, and we hope that the release of this implementation could encourage further research into fast algorithms for kernel density estimation in high dimensions.

\bibliographystyle{plain}
\bibliography{reference}

\appendix

\section{User Guide Examples using STAG Python} \label{app:ug_python}
This section includes example code omitted from Section~\ref{sec:user_guide}
demonstrating how to use STAG Python for locality sensitive hashing, kernel density estimation, and spectral clustering.
The online documentation\footnote{\url{https://staglibrary.io/docs/python/}} contains the full API description of the library.

\subsection{Handling Data}
The following example shows how to read and write data to a CSV file.
\begin{lstlisting}[language=Python]
  import stag.data

  # Read data from disk
  data = stag.data.load_matrix("data.csv")

  # Write data to disk
  stag.data.save_matrix(data, "my_data.csv")
\end{lstlisting}
STAG Python uses the \texttt{stag.utility.DenseMat} class to represent a data matrix.
The \texttt{to\_numpy()} method converts the object to a numpy matrix.
\begin{lstlisting}[language=Python]
  import numpy
  import stag.data

  # Read data from disk and convert to a numpy matrix
  data = stag.data.load_matrix("data.csv")
  data_np = data.to_numpy()
\end{lstlisting}
The following example shows how to create a \texttt{DataPoint} object representing one row of the data matrix.
\begin{lstlisting}[language=Python]
  import stag.data

  # Create a data point referencing the 10th row of the data matrix
  data = stag.data.load_matrix("data.csv")
  point = stag.data.DataPoint(data, 10)
\end{lstlisting}

\subsection{Locality Sensitive Hashing}
STAG Python includes the \texttt{stag.lsh.LSHFunction} and \texttt{stag.lsh.E2LSH} classes with the same functionality as the C++ library.
The following code creates and applies an LSH function.
\begin{lstlisting}[language=Python]
  import stag.data
  import stag.lsh

  # Create an LSH function for 10-dimensional data.
  f = stag.lsh.LSHFunction(10)

  # Apply the hash function to a data point
  point = stag.data.DataPoint(data, 10)
  h = f.apply(point)

  # Compute the collision probability for points at distance c
  c = 1
  prob = stag.lsh.LSHFunction.collision_probability(c)
\end{lstlisting}
The following example creates a Euclidean LSH hash table for a dataset.
\begin{lstlisting}[language=Python]
  import stag.data
  import stag.lsh

  # Load the data from disk
  data = stag.data.load_matrix("data.csv")

  # Create the Euclidean LSH table
  K = 6
  L = 50
  dps = [stag.data.DataPoint(data, i) for i in range(data.rows())]
  table = stag.lsh.E2LSH(K, L, dps)

  # Find the data points close to a query point
  q = stag.data.DataPoint(data, 10)
  close_points = table.get_near_neighbours(q)
\end{lstlisting}

\subsection{Kernel Density Estimation}
STAG Python provides methods for kernel density estimation in the \texttt{stag.kde} module.
The Gaussian kernel similarity between two points is computed in the next example.
\begin{lstlisting}[language=Python]
  import stag.data
  import stag.kde

  # Load the data matrix
  data = stag.data.load_matrix("data.csv")
  p1 = stag.data.DataPoint(data, 1)
  p2 = stag.data.DataPoint(data, 2)

  # Compute the Gaussian kernel similarity between the points
  a = 1
  val = stag.kde.gaussian_kernel(a, p1, p2)
\end{lstlisting}
The following example demonstrates the \texttt{stag.kde.ExactGaussianKDE} class.
\begin{lstlisting}[language=Python]
  import stag.data
  import stag.kde

  # Load the data matrix
  data = stag.data.load_matrix("data.csv")

  # Create the kernel density data structure
  a = 1
  kde = stag.kde.ExactGaussianKDE(data, a)

  # Compute the kernel density for a query point
  q = stag.data.DataPoint(data, 10)
  kernel_density = kde.query(q)

  # Compute the kernel densities for many query points
  kernel_densities = kde.query(data)
\end{lstlisting}
STAG Python also includes the \texttt{stag.kde.CKNSGaussianKDE} class with the same constructors as described in Section~\ref{sec:ug_kde}.
\begin{lstlisting}[language=Python]
  import stag.data
  import stag.kde

  # Load the dataset
  data = stag.data.load_matrix("data.csv")

  # Create the CKNS data structure using the simple constructor
  a = 1
  kde = stag.kde.CKNSGaussianKDE(data, a)

  # Specify the values of eps and min_mu
  eps = 0.5
  min_mu = 0.0001
  kde = stag.kde.CKNSGaussianKDE(data, a, eps=eps, min_mu=min_mu)

  # Fully specify the constants used within the data structure
  k1 = 10
  k2_constant = 5
  p_offset = 1
  kde = stag.kde.CKNSGaussianKDE(data, a, min_mu=min_mu, k1=k1, k2_constant=k2_constant, sampling_offset=p_offset)
\end{lstlisting}
The kernel density estimate for a query point can be computed with the \texttt{query} method.
\begin{lstlisting}[language=Python]
  # Query a single data point
  q = stag.data.DataPoint(data, 10)
  kd_estimate = kde.query(q)

  # Compute estimates for many data points
  kd_estimates = kde.query(data)
\end{lstlisting}

\subsection{Fast Spectral Clustering}
The following example demonstrates how to construct an approximate similarity graph and perform spectral clustering with STAG Python.
\begin{lstlisting}[language=Python]
  import stag.data
  import stag.graph
  import stag.cluster

  # Load the dataset
  data = stag.data.load_matrix("data.csv")

  # Create the approximate similarity graph
  a = 1
  g = stag.cluster.approximate_similarity_graph(data, a)

  # Find 10 clusters in the graph
  k = 10
  labels = stag.spectral_cluster(g, k)
\end{lstlisting}

\end{document}